\documentclass[manuscript]{aastex}

\usepackage{graphicx}

\shorttitle{Chemical Abundances of 2MASS\,GC02 and Mercer~5}
\shortauthors{Pe\~naloza et al.}

\begin{document}
\title{Chemical Abundances of the Highly Obscured Galactic Globular Clusters 2MASS\,GC02 and Mercer~5}
\author{Francisco Pe\~naloza}
\affil{Insituto de Fisica y Astronomia, Facultad de Ciencias, Universidad de Valparaiso}
\affil{Av. Gran Bretana 1111, Valparaiso, Chile}
\affil{francisco.penaloza@uv.cl}
\author{Peter Pessev\altaffilmark{1}}
\affil{Instituto de Astrof\'isica de Canarias (IAC); Gran Telescopio Canarias (GRANTECAN) }
\affil{E-38205 La Laguna, Tenerife, Spain}
\affil{peter.pessev@gtc.iac.es}
\author{Sergio Va\'squez}
\affil{Instituto de Astrof\'isica, Pontifica Universidad Cat\'olica de Chile; Milenium Institute of Astrophysics, MAS}
\affil{Av. Vicu\~na Mackenna 4860, Santiago, Chile}
\affil{svasquez@astro.puc.cl}
\author{Jura Borissova, Radostin Kurtev}
\affil{Insituto de Fisica y Astronomia, Facultad de Ciencias, Universidad de Valparaiso; Milenium Institute of Astrophysics, MAS}
\affil{Av. Gran Bretana 1111, Valparaiso, Chile}
\affil{jura.borissova@uv.cl, radostin.kurtev@uv.cl}
\and
\author{Manuela Zoccali}
\affil{Instituto de Astrof\'isica, Pontifica Universidad Cat\'olica de Chile; Milenium Institute of Astrophysics, MAS}
\affil{Av. Vicu\~na Mackenna 4860, Santiago, Chile}
\affil{mzoccali@astro.puc.cl}
\slugcomment{Submitted to PASP}

\begin{abstract}

We present the first high spectral resolution abundance analysis of two newly discovered Galactic globular clusters, namely Mercer~5 and 2MASS\,GC02 residing in regions of high interstellar reddening in the direction of the Galactic center. The data were acquired with the Phoenix high-resolution near-infrared echelle spectrograph at Gemini South (R $\sim 50000$) in the 15500.0 - 15575.0 spectral region. Iron, Oxygen, Silicon, Titanium and Nickel abundances were derived for two red giant stars, in each cluster, by comparing the entire observed spectrum with a grid of synthetic spectra generated with MOOG. We found [Fe/H] values of $-0.86 \pm 0.12$ and $-1.08 \pm 0.13$ for Mercer~5 and 2MASS\,GC02 respectively.
The  [O/Fe], [Si/Fe] and [Ti/Fe] ratios  of the measured stars of Mercer~5 follow the general 
trend of both bulge field and cluster stars at this metallicity, and are enhanced by $\geq +$ 0.3. The 2MASS\,GC02 stars have relatively  lower ratios, but still compatible with other bulge clusters. Based on metallicity and abundance patterns of both objects we conclude that these are typical bulge globular clusters.

\end{abstract}


\section{Introduction}

The properties of globular clusters and of their stellar populations provide fundamental information on the environment where galaxies formed, on the Galactic formation process, and are a basic ingredient for the understanding of the stellar populations in the external galaxies. Moreover, the properties of globular clusters are deeply connected with the history of their host galaxy. We believe today that galaxy collisions, galaxy cannibalism, as well as galaxy mergers leave their imprint on the globular cluster population of any given galaxy. Thus, when investigating globular clusters we hope to be able to use them as an acid test for our understanding of the formation and evolution of galaxies. The Galactic bulge is particularly important in the context of galaxy formation, as it is the only bulge that can be resolved into stars down to the bottom of the main sequence, and for which chemical abundances can be obtained with high-resolution spectra. Determinations of detailed chemical compositions are key data for studies of the origin and evolution of stellar populations, since they carry characteristic signatures of the objects that enrich the interstellar gas. The Galactic bulge globular clusters are relatively poorly understood stellar systems. The number of bulge globular cluster stars for which detailed chemical abundance information is available is considerably smaller compared to stars in halo clusters. Moreover, the advent of the new generation extensive surveys such as SDSS \citep{abazajian_et_al_2009}, 2MASS \citep{skrutskie_et_al_2006}, GLIMPSE \citep{benjamin_et_al_2003}, VISTA Variables in the Via Lactea (VVV) Public Survey \citep{minniti_et_al_2010, saito_et_al_2012} yielded detection of several new Galactic Globular Clusters (GGCs). The December 2010 compilation of the \citet{harris_1996} catalog included seven new GGCs not present in  the February 2003 version, but several more cluster candidates have been proposed in the last years: SDSSJ1257+3419 \citep{sakamoto_hasewaga_2006}, FSR~584 \citep{bica_et_al_2007}, FSR~1767 \citep{bonatto_et_al_2007}, FSR~190 \citep{froebrich_et_al_2008a}, Pfleiderer~2 \citep{ortolani_et_al_2009}, VVV\,CL001 \citep{minniti_et_al_2011}, Mercer~5 \citep{longmore_et_al_2011}, VVV\,CL002 \citep{moni_bidin_et_al_2011} and Kronberger~49 \citep{ortolani_et_al_2012}. Thus, detailed investigation of these newly discovered members of the globular cluster family can contribute significantly to the global understanding of the whole system. In this study we report the results of our high-resolution Phoenix spectroscopy of selected red giant stars of two newly-discovered globular clusters: 2MASS\,GC02 and Mercer~5, projected in the bulge area of the Galaxy. The globular cluster 2MASS\,GC02 was reported by \citet{hurt_et_al_2000} and was detected within the Two Micron All Sky Survey (2MASS). Later on \citet{borissova_et_al_2007} obtained deep infrared images and low-resolution K-band spectra. Based on the analysis of the J-Ks versus Ks color-magnitude diagram and spectroscopically derived metallicities and radial velocities of 15 stars they concluded that the cluster is moderately metal-rich ([Fe/H]=-1.1) and has a relatively high radial velocity. Its horizontal branch appears to be predominantly red, though the photometry can not rule out presence of a blue component as seen in NGC\,6388 and NGC\,6441. Comparison with the existing kinematic and abundance information for the GGCs indicates that 2MASS\,GC02 most probably belongs to the bulge sub-population, although inner halo association can not be ruled out. Recently, \citet{alonso_garcia_et_al_2014} discovered 29 new variables inside the tidal radius of 2MASS\,GC02, using the Vista varaibles in the Via Lactea (VVV) ESO Large Public Survey. Eight of these variables are classified as RR Lyr stars. Using these newly discovered RR Lyrae stars, they found that the extinction towards the cluster is highly differential, and seems to follow a non-standard law, thus putting the cluster closer to the galactic center (calculated distance of RGc = 2.2 Kpc).

The dust-obscured Galactic star cluster Mercer~5 was investigated by \citet{longmore_et_al_2011}. The analysis of the near-infrared photometry from the United Kingdom Infrared Digital Sky Survey (UKIDSS) and the SofI/NTT near-IR spectroscopy, indicate that the object almost certainly is a Galactic Globular Cluster, located at the edge of the Galactic bulge. The cluster suffers strong and variable extinction, located at a distance of approximately 5.5 kpc and is also moderately metal-rich ([Fe/H]=-1.0).

\section{Data, Reduction and Analysis}

\begin{figure}
\epsscale{1.00}
\plotone{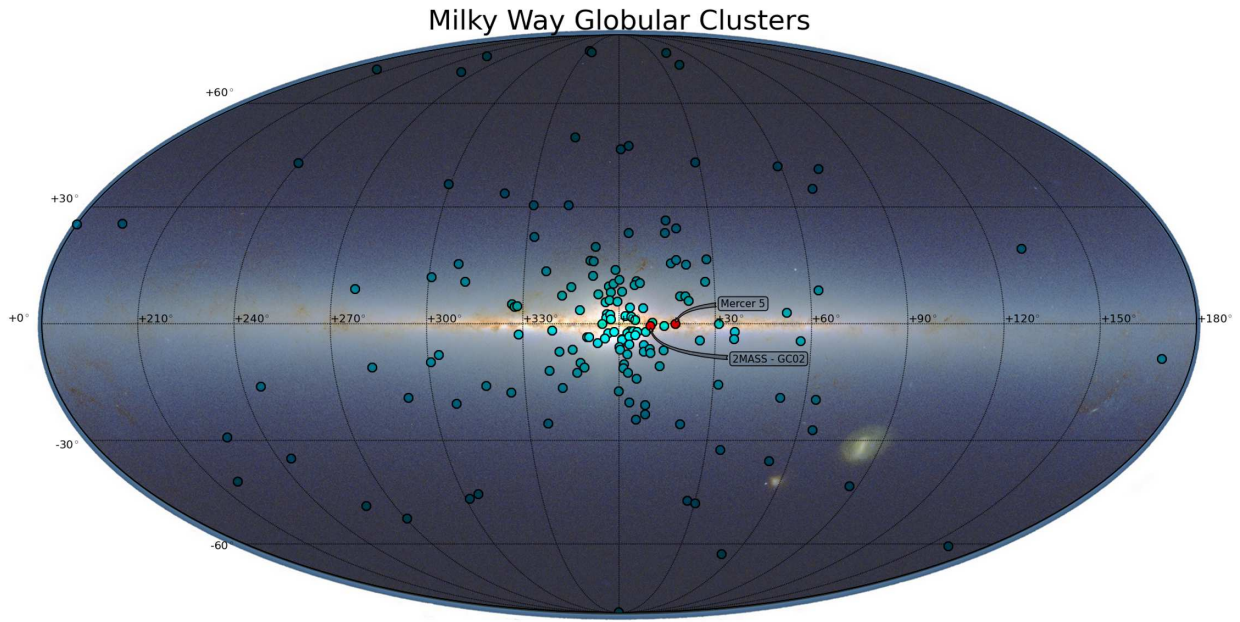}
\caption{Illustration of the spatial distribution of the Milky Way Globular Clusters around the plane of the Galaxy. The two clusters targeted by the current study are marked and labeled. An elliptical projection of the Galactic coordinates was used with an all-sky image of the 2MASS stellar density set as a background. 
\label{fig:fig1}}
\end{figure}

Relevant information about our target clusters is presented in Table~\ref{tab:tab1}. Note that Mercer~5 is a newly discovered cluster \citep{mercer_et_al_2005}, still not included in the online version of the \citet{harris_1996} catalog (2010 edition). Both globular clusters targeted for observations are located at low Galactic latitude, close to the plane of the Milky Way, in regions of high interstellar reddening (see Figure~\ref{fig:fig1}). Hence Phoenix high-resolution near-infrared echelle spectrograph \citep{hinkle_et_al_1998} mounted at Gemini South 8-m telescope was a natural choice for the observations. The combination of large telescope aperture and high spectral resolution is crucial for accurate abundance determination, considering the apparent magnitudes of the individual red giants in our sample. More specifically, the data reported in Table~\ref{tab:tab1} are taken from \citet{longmore_et_al_2011} (Mercer~5) and \citet{borissova_et_al_2007} (2MASS\,GC02). The fundamental parameters of both clusters are calculated using the technique outlined in \citet{Ferraro_et_al_2006}; \citet{Valenti_et_al_2005} and \citet{Valenti_et_al_2007}, which allows to determine the reddening, distance modulus, and a global photometric metallicity of a globular cluster from its near-infrared CMD. In this case the RGB and HB clump calibrations were used. The targeted wavelength range was selected based on the line list published by \citet{ryde_et_al_2010} and covers a variety of Iron and $\alpha$--elements metal lines. It also has the advantage of being devoid of bright OH airglow lines, which aids the analysis of faint spectral features. The Phoenix configuration that was used is presented in Table~\ref{tab:tab2}. Note that the spectral coverage provided by Phoenix is limited by the size of the science array and is much smaller than the bandwidth of the H6420 order-sorting filter.

\begin{table}
\begin{center}
\caption{Information about the targeted clusters \label{tab:tab1}}
\begin{tabular}{lllrlcrcl}
\tableline
ID & \multicolumn{1}{c}{RA} & \multicolumn{1}{c}{DEC} & \multicolumn{1}{c}{l} & \multicolumn{1}{c}{b} & \multicolumn{1}{c}{D} & \multicolumn{1}{c}{$E(J-K)$} &  \multicolumn{1}{c}{$(m-M)_{0}$} & Ref. \\
                   & \multicolumn{1}{c}{hh:mm:ss} & \multicolumn{1}{c}{dd:mm:ss}  & \multicolumn{1}{c}{deg.} & \multicolumn{1}{c}{deg.} & \multicolumn{1}{c}{arcmin.} & \multicolumn{1}{c}{mag.} & \multicolumn{1}{c}{mag.} & \\[5pt]
(1)& \multicolumn{1}{c}{(2)} & \multicolumn{1}{c}{(3)}  & \multicolumn{1}{c}{(4)} & \multicolumn{1}{c}{(5)} & \multicolumn{1}{c}{(6)} & \multicolumn{1}{c}{(7)} & \multicolumn{1}{c}{(8)} & (9)\\[5pt]
\tableline
2M\,GC02 & 18:09:37 & -20:46:44 &  9.78 & -0.62 & $1.9\pm0.2$ &  $3.0\pm0.1$ & $13.54\pm0.15$ \ \ & $a,b$ \\[5pt]
Mercer~5 & 18:23:19 & -13:40:02 & 17.59 & -0.11 & 0.60        &  $2.1\pm0.7$ & $14.29^{(1)}$ & $c,d$ \\[5pt]
\hline
\multicolumn{9}{l}{\ \ \ \ \ \ Notes: Column (1) is the cluster ID, followed by the equatorial coordinates of the ob-}\\
\multicolumn{9}{l}{\ \ \ \ \ \ \ \ \ \ \ \ \ \ \ \  ject (columns (2) and (3)). The Galactic coordinates are presented in columns}\\
\multicolumn{9}{l}{\ \ \ \ \ \ \ \ \ \ \ \ \ \ \ \  (4) and (5). Column (6) shows the apparent diameter of the cluster, followed by}\\
\multicolumn{9}{l}{\ \ \ \ \ \ \ \ \ \ \ \ \ \ \ \  an estimate of the color excess E(J-K) in column (7). The distance mo-}\\
\multicolumn{9}{l}{\ \ \ \ \ \ \ \ \ \ \ \ \ \ \ \  dulus to the object is listed in column (8), followed by the list of the references}\\
\multicolumn{9}{l}{\ \ \ \ \ \ \ \ \ \ \ \ \ \ \ \  to the various sources of information used in the table.}\\
\multicolumn{9}{l}{ \ \ \ \ \ \ \ \ \ \ \ \ \ \ \ \  (1) The maximum $(m-M)_{0}$ value from \citet{longmore_et_al_2011} is considered.}\\
\multicolumn{9}{l}{References: $(a)$ \citet{borissova_et_al_2007}, $(b)$ \citet{hurt_et_al_2000}, $(c)$ \citet{mercer_et_al_2005},}\\
\multicolumn{9}{l}{ \ \ \ \ \ \ \ \ \ \ \ \ \ \ \ \ $(d)$ \citet{longmore_et_al_2011}}\\
\end{tabular}
\end{center}
\end{table}

\subsection{Red Giant Stars Sample Selection and Observations}

\begin{figure}[htb]
\centering
    \includegraphics[width=.33\textwidth]{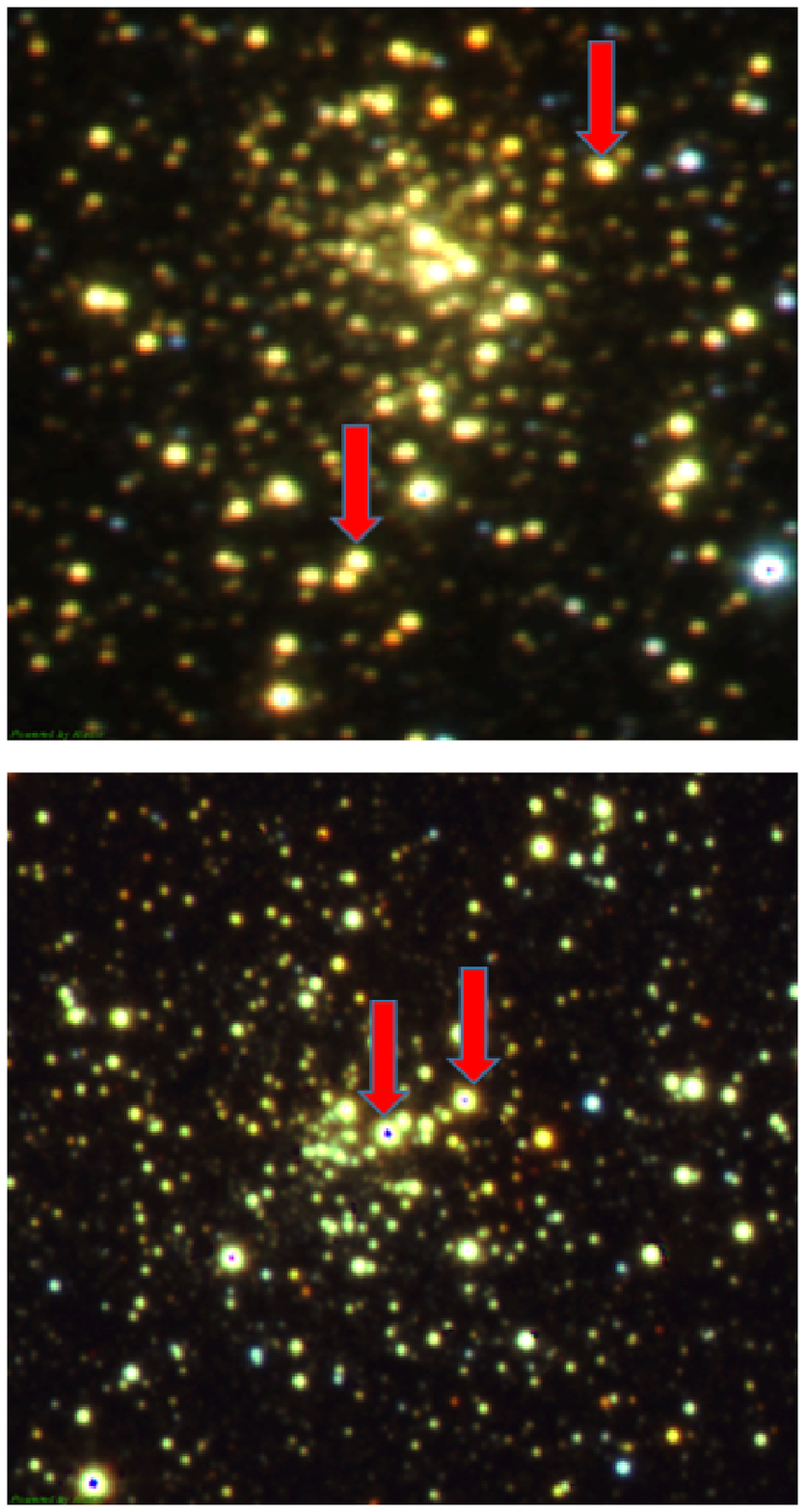}\hfill
    \includegraphics[width=.46\textwidth]{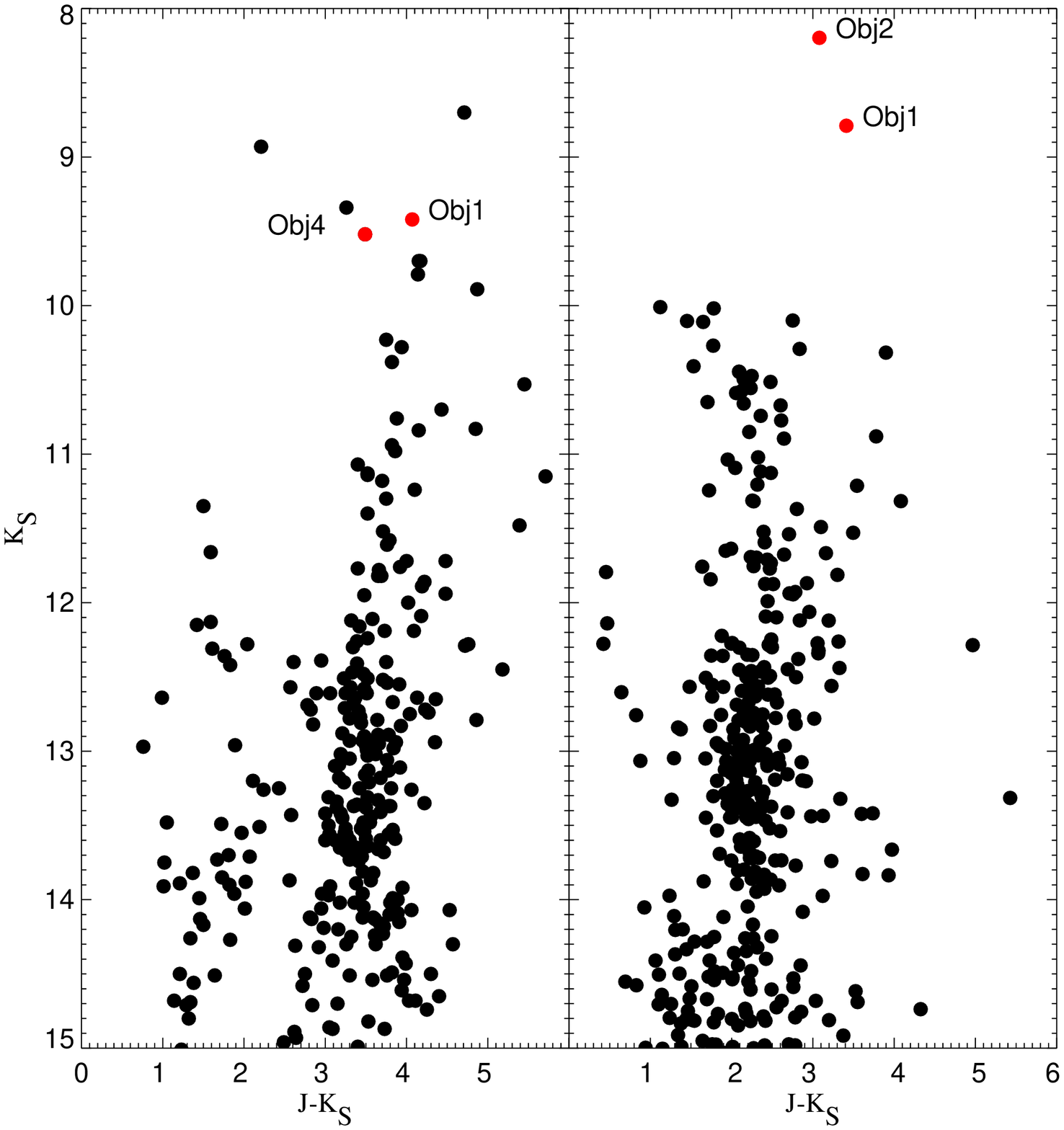}\hfill
\caption{Illustration of the spatial distribution of the targeted red giants into the clusters and their positions on the CMDs. Left panel shows $1'$ x $1'$ three color images of the fields of GC02 and Mercer~5, taken from Vista variables in the Via Lactea (VVV) and UKIDS Galactic Plane Surveys. Right: The color-magnitude diagrams of the GC02 and  Mercer~5 cluster fields. The black points represent all objects residing into a 60\arcsec radius; the observed objects are labeled. 
\label{fig:fig2}}
\end{figure}

The stars observed in each cluster were selected on the basis of pre-existing Near-IR CMDs and low-resolution spectroscopy (SofI/NTT, R$\sim$1500 and ISAAC @ VLT, R$\sim$500). All of them were identified as high probability members of their host star clusters. The information about the individual stars observed is compiled in Table~\ref{tab:tab3}. We observed stars close to the Tip of the Red Giant Branch (TRGB) in order to ensure that the spectra will reach the required S/N ($\sim  60$) with a sensible investment of observing  time. Figure~\ref{fig:fig2} shows the positions of the target stars in the clusters, and on their color-magnitude diagrams. The J, H and Ks images are taken from the Vista variables in the Via Lactea Survey (VVV, DR2, http://horus.roe.ac.uk/vsa/, \citet{saito_et_al_2012}) and UKIDS Galactic Plane Surveys (GPS,DR7, http://www.ukidss.org/index.html) and the three-color images are constructed. The CMDs of the clusters are build using the photometric catalogs provided in VVV and GPS and include all objects residing into a 60\arcsec \, radius.  
 
The spectra were acquired at Gemini South 8-m telescope during the semester 2010A (Program GS-2010A-Q-30, PI P.Pessev). Since our targets are relatively faint for such high spectral resolution (H$\sim$10-11), our observations took full advantage of the queue mode of operation, allowing us to impose exactly the required sky and seeing conditions during the data acquisition. A standard technique of ABBA offset pattern across the slit was used. Due to the crowded nature of the observed fields (see Figure~\ref{fig:fig2}) we did a quick pre-imaging for the Phoenix acquisitions and provided detailed finder charts for the queue observers. Telluric standards, of spectral class A or earlier, from the Gemini calibration library were observed either before or/and after the observations at matching airmass to ensure proper reduction and calibration. Since standard stars are significantly brighter than the science targets a larger offset along the slit (4\arcsec) was used, with respect to that of the science data (2.5\arcsec). According to the standard Gemini procedure, the exposure times for the tellurics were adjusted by the night observer (depending on the luminosity of the particular star and the observing conditions at the moment of the observation) to provide sufficient S/N for high quality calibration. In general the exposure times for the standards were much shorter with respect to those of the science targets. Flat fields were taken each time science data were acquired, before moving the grating or changing the instrument configuration, using the dedicated 100W GCAL calibration source. Phoenix darks with exposure times matching the flat field data were secured at the end of each night. The calibration dataset was completed by wavelength calibration frames acquired with the internal Phoenix ThAr lamp. Considering the significant investment of observing time required and taking into account the narrow wavelength coverage of Phoenix (that does not provide a favorable configuration of calibration lines on the detector), these were taken only for a fraction of the data as an extra wavelength reference cross-check.

\begin{table}
\begin{center}
\begin{tabular}{llrl}
\tableline
Slit width & & 4 & [pix.]\\ 
Slit width & &0.34 & [arcsec.]\\ 
Resolution ($\lambda$/$\Delta\lambda$) & & $\sim$ 50000 &\\
\hline
Central wavelength & & $\sim$ 15537.5 & [\AA]\\
Wavelength coverage & min. & $\sim$ 15500.0 & [\AA]\\
& max. & $\sim$ 15575.0 & [\AA]\\
Width of the observed spectral region & & 75 & [\AA]\\
\hline
Filter used & & H6420 & \\
\tableline
\end{tabular}
\caption{Phoenix configuration used during the observations 
\label{tab:tab2}}
\end{center}
\end{table}

\begin{table}
\begin{center}
\caption{Observing log and information about the individual targeted stars \label{tab:tab3}}
\begin{tabular}{llllrrcc}
\tableline
Cluster ID & StarID & \multicolumn{1}{c}{RA} & \multicolumn{1}{c}{DEC} & \multicolumn{1}{l}{D cen.} & \multicolumn{1}{c}{H} & \multicolumn{1}{c}{Date Obs.} & \multicolumn{1}{c}{Exp. Time}\\
            &       & \multicolumn{1}{c}{hh:mm:ss} & \multicolumn{1}{c}{dd:mm:ss}  & \multicolumn{1}{c}{arcsec.} & \multicolumn{1}{c}{mag.} & \multicolumn{1}{c}{DDMMYYYY} & \multicolumn{1}{c}{sec.} \\
 (1) & (2) & \multicolumn{1}{c}{(3)} & \multicolumn{1}{c}{(4)}  & \multicolumn{1}{c}{(5)} & \multicolumn{1}{c}{(6)} & \multicolumn{1}{c}{(7)} & \multicolumn{1}{c}{(8)}\\
\tableline
2M\,GC02 & Star 1 & 18:09:35.20 & -20:47:02.20 &  32.56  & 10.7 & 04062010UT & 7200\\
2M\,GC02 & Star 4 & 18:09:36.50 & -20:46:44.10 & 7.50 & 9.7 & 07062010UT & 7200\\[7pt]
Mercer~5 &  Star 1 & 18:23:19.08 & -13:40:09.90 & 8.05 & 9.9 & 02072010UT & 3000\\
Mercer~5 & Star 2 & 18:23:19.58 & -13:40:06.70 &  10.04 & 9.1 & 07062010UT & 2400\\
\hline
\multicolumn{8}{l}{Notes: Columns (1) and (2) are respectively the cluster and star ID, followed by the equa-}\\
\multicolumn{8}{l}{torial coordinates of the object (columns (3) and (4)). Column (5) is showing the distance}\\
\multicolumn{8}{l}{between the individual star and the center of the corresponding cluster. The H magnitude}\\
\multicolumn{8}{l}{of the object is given in column (6). The last two columns (7) and (8) are listing the UT}\\
\multicolumn{8}{l}{date of the observation and the total exposure time used for the observations.}\\
\end{tabular}
\end{center}
\end{table}

\subsection{Data Reduction}

\begin{figure}
\centering
    \includegraphics[width=0.85\textwidth]{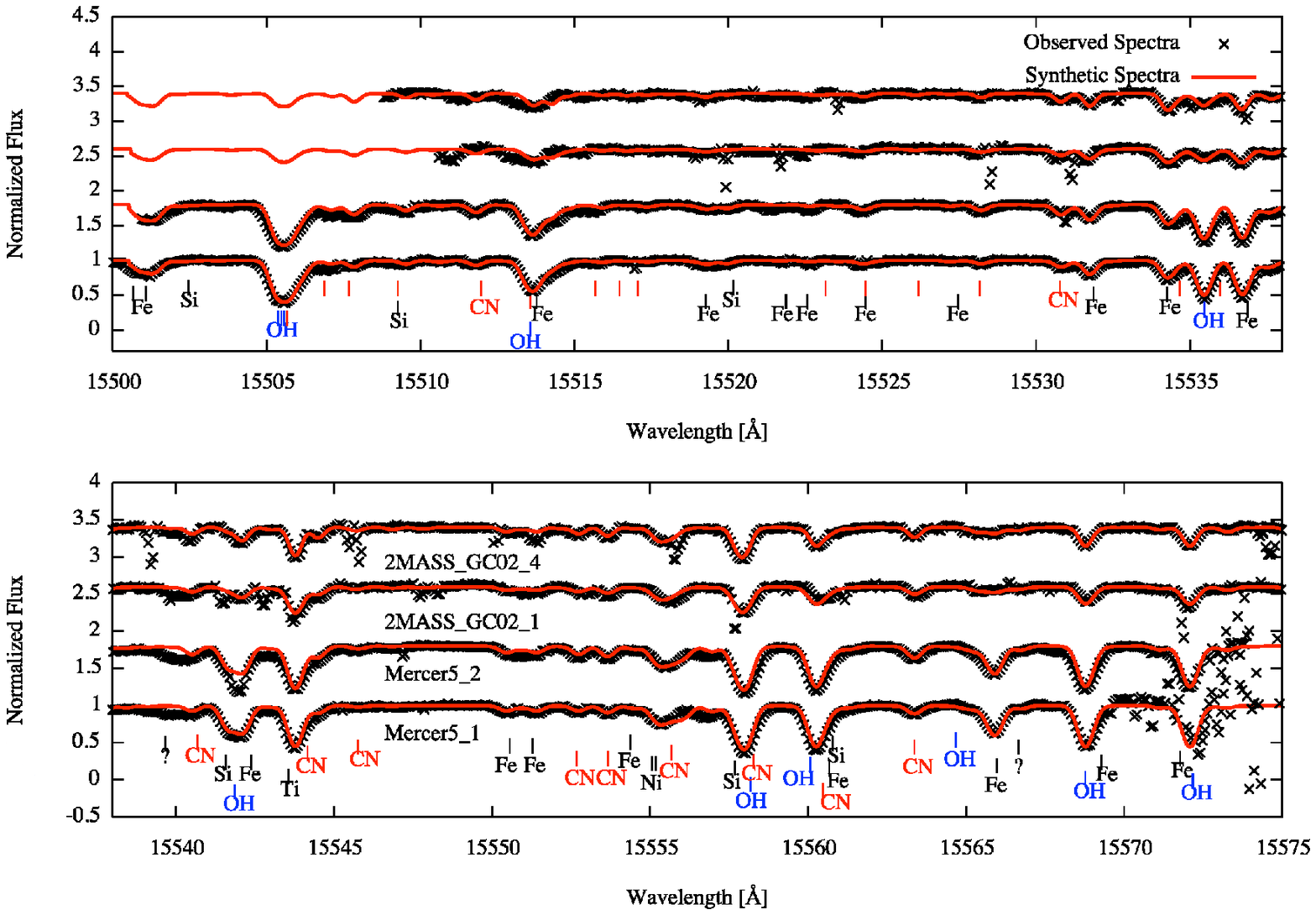}\\
\caption{Observed spectra are shown with black crosses. Our best synthetic spectra are shown with a red line. Metal and Molecular lines are indicated below the spectra.}
\label{fig:fig3}
\end{figure}

The reduction of the spectra was carried out in the IRAF{\footnote{IRAF is distributed by the National Optical Astronomy Observatory, which is operated by the Association of Universities for Research in Astronomy (AURA) under cooperative agreement with the National Science Foundation.}} environment, using the standard procedure for Phoenix data{\footnote{Available at: ftp://iraf.noao.edu/iraf/misc/phoenix.readme}}. Here we provide only a brief outline, with a focus on some crucial steps. First we need to trim all the science and calibration frames. This is important because a small section of the detector array is delaminated and the first $\sim$ 50 rows of each image are infested with a lot of bad pixels. Skipping that step will cause unnecessary complications during the entire data reduction process. Further the flats and darks associated with each set of observations were combined and subtracted from the combined flats. The step of developing the normalized master flat for each set is particularly important, in order to properly remove the features due to variations in the slit illumination. Normalization is also crucial for the reduction of the telluric standards, to avoid spurious features affecting the final results. OH airglow lines were removed from both standard and science targets stellar spectra by subtracting each one of the ABBA pairs. Then the two-dimensional frames were divided by the normalized flat fields before the extraction of the individual spectra. The one-dimensional spectra were wavelength calibrated, using atmospheric OH airglow lines. We targeted a spectral region that contains multiple interesting lines of Iron and $\alpha$--process elements being devoid of bright airglow features. This is particularly beneficial for the data reduction and the analysis of weak spectral lines, but poses significant challenges for the wavelength calibration. Most of the available atlases of the OH airglow are not suitable for analysis of such high-resolution data, especially taking into account the width of the analyzed spectral region (see Table~\ref{tab:tab2}). Fortunately the long exposures on the science targets allowed to identify five airglow features and assign the corresponding wavelengths using the The Arcturus Atlas (telluric lines) obtained with Phoenix at Kitt Peak{\footnote{Available online at: ftp://cdsarc.u-strasbg.fr/cats/J/PASP/107/1042/}} \citep{hinkle_et_al_1995}. The wavelength calibration solution was then cross-checked against the obtained arc lamp exposures. The telluric features in the final one-dimensional spectrum were corrected using the data for the corresponding standard stars. The resulting spectra for each of the stars in both 2MASS\,GC02 and Mercer~5 are presented on Figure~\ref{fig:fig3}.

\subsection{Analysis of the Obtained Spectra}

Stellar spectra and chemical abundances were analyzed using the MOOG code \citep{sneden_1973}. To perform the analysis the program relies on stellar atmospheres models and line lists of the atomic and molecular species in the studied wavelength range. The model atmospheres were computed with the ATLAS9 \citet{kurucz_1993} code. Recently a significant progress was achieved on the availability of the high resolution near-IR line lists, but it is still much more limited, compared to the optical wavelength range. Pioneering works of \citet{wallace_et_al_1996} and \citet{hinkle_et_al_1995} focused on hight-resolution, high signal-to-noise spectra of the Sun and Arcturus. Current effort is aimed on provide more uniform coverage across the HR diagram (see \citet{lebzelter_et_al_2012}). Although this is a massive improvement over the earlier situation, more data and further studies are needed to match the optical spectral atlases for reference stars. The line list we used was kindly provided by Nills Ryde (private communication). It covers 699 atomic and molecular lines in the 15500 - 15575 A wavelength region (see Table~6), including Iron lines, lines of $\alpha$--elements and molecular lines (CN and OH).

MOOG computes synthetic spectra based input parameters, such as effective temperature, surface gravity, metallicity, micro-turbulence and $\alpha$-elements abundance. In order to determine the effective temperature we used the photometry and low-resolution spectroscopy published by \citet{borissova_et_al_2002} and \citet{borissova_et_al_2007} for 2MASS\,GC02 and \citet{longmore_et_al_2011} for Mercer~5 in conjunction with the $T_{eff}$:color:[Fe/H] calibrations of \citet{ramirez_melendez_2005} for giant stars. The initial effective temperatures used were 4000K for the 2MASS\,GC02 stars and 3600K for Mercer~5. Surface gravity ($log\  g$) has been estimated from theoretical evolutionary tracks using the location of the stars on the red giant branch \citet{origlia_et_al_1997}. We performed spectral synthesis on suitable Fe \textsc{I} and Fe \textsc{II} lines to derive the metallicity. The micro-turbulence velocity is set to a value typical for red giant stars ($\xi_{micro}$[km $s^{-1}$] = 2.13 - 0.23 log g) as given in \citet{kirby_et_al_2009}. The adopted [$\alpha/Fe$]$= 0.40$ is set in accordance with \citet{gonzalez_et_al_2011}. Their results are based on the analysis of 650 giants in the different sections of the Galactic Bulge. To fit the widths and shapes of the lines of the observed spectra, each synthetic spectrum was convolved with a gaussian function and macro-turbulence function  and the abundance is allowed to vary until the best fit is identified. The selected spectral range gives a reasonable number of atomic and molecular lines not affected by blending to derive relative abundances (see Figure~\ref{fig:fig3}). Unfortunately it does not cover CO molecular lines. Therefore, we adopted three values of $[C/Fe]$ = -0.15, -0.35, and -0.55 in accordance with the results of \citet{ryde_et_al_2009} and \citet{ryde_et_al_2010}. The total abundance errors were determined by varying each of the input stellar parameters by its estimated errors and adding in quadrature the resulting abundance variations (see Table~\ref{tab:tab4}). We estimate the typical uncertainties of $\Delta T_{eff}  \pm$ 100K, $\Delta$ log g $\pm$ 0.2 dex, $\Delta \xi_{micro} \pm$ 0.5  km  $s^{-1}$, which translates into abundance errors for Fe$\sim$0.14, N$\sim$0.10, O$\sim$0.11, Si$\sim$0.14, Ti$\sim$0.17, Ni$\sim$0.17, respectively.

\begin{table}[h]
\caption{Stellar parameters and derived abundances}
\begin{tabular}{l*{6}{c}r}
\hline 
                           &  & Mercer~5 &     &  &  2MASS\,GC02 &    \\
\hline
Object:                   &  & \#1  & \#2  &  & \#1 & \#4  \\

$T_{eff}$ (K)              &  & 3650$\pm$100 & 3680$\pm$100 &  &  4000$\pm$100   &  4050$\pm$100  \\

Log g $(cm\,s^{-2})$       &  & 0.5$\pm$0.2 & 0.5$\pm$0.2 &   &  1.0$\pm$0.2 & 1.0$\pm$0.2  \\

$\xi_{micro}$ $(km\,s^{-2})$&  & 2.0$\pm$0.5 & 2.0$\pm$0.5 &   &  1.9$\pm$0.5 & 1.9$\pm$0.5  \\

$[\alpha/Fe]$ (dex)     &  & 0.40 & 0.40 &  & 0.40  & 0.40 \\

$[Fe/H]$ (dex)         &  & -0.90$\pm$0.13 & -0.80$\pm$0.14 & &  -1.10$\pm$0.14 &  -1.05$\pm$0.14 \\

\hline

                        & & & $[C/Fe]$ = -0.15 dex & & &\\
\hline

$[N/Fe]$ (dex)         &  & +0.45$\pm$0.09 & +0.65$\pm$0.10 &  & +0.53$\pm$0.10  & +0.43$\pm$0.10 \\

$[O/Fe]$ (dex)         &  & +0.30$\pm$0.11 & +0.30$\pm$0.11 &  & +0.12$\pm$0.12  & +0.33$\pm$0.12  \\

$[Si/Fe]$ (dex)         &  & +0.50$\pm$0.14 & +0.55$\pm$0.14 &  & +0.03$\pm$0.15 &   0.00$\pm$0.15\\

$[Ti/Fe]$ (dex)        &  & +0.30$\pm$0.17 & +0.48$\pm$0.17 &  & +0.20$\pm$0.17  &  +0.35$\pm$0.17\\

$[Ni/Fe]$ (dex)        &  & +0.30$\pm$0.17 & +0.25$\pm$0.17 &  & +0.20$\pm$0.17  &  +0.10$\pm$0.17 \\
\hline
                        & & & $[C/Fe]$ = -0.35 dex & & &\\
\hline

$[N/Fe]$ (dex)         &  & +0.65$\pm$0.10 & +0.95$\pm$0.10 &  & +0.75$\pm$0.10   & +0.70$\pm$0.10  \\

$[O/Fe]$ (dex)         &  & +0.30$\pm$0.10 & +0.32$\pm$0.11 &  & +0.10$\pm$0.12   & +0.35$\pm$0.12  \\

$[Si/Fe]$ (dex)         &  & +0.50$\pm$0.14 & +0.55$\pm$0.14 &  & +0.05$\pm$0.15 & +0.02$\pm$0.15  \\

$[Ti/Fe]$ (dex)        &  & +0.40$\pm$0.17 & +0.48$\pm$0.17 &  & +0.30$\pm$0.17  & +0.35$\pm$0.17 \\

$[Ni/Fe]$ (dex)        &  & +0.30$\pm$0.17 & +0.25$\pm$0.17 &  & +0.15$\pm$0.17  & +0.10$\pm$0.17  \\

\hline
                        & & & $[C/Fe]$ = -0.55 dex & & &\\
\hline

$[N/Fe]$ (dex)         &  & +0.90$\pm$0.10 & +1.15$\pm$0.11 &  & +1.03$\pm$0.10  & +0.90$\pm$0.10  \\

$[O/Fe]$ (dex)         &  & +0.30$\pm$0.10 & +0.32$\pm$0.11 &  & +0.15$\pm$0.11  & +0.30$\pm$0.11  \\

$[Si/Fe]$ (dex)         &  & +0.50$\pm$0.14 & +0.55$\pm$0.14 &  & +0.03$\pm$0.15 & 0.00$\pm$0.15  \\

$[Ti/Fe]$ (dex)        &  & +0.40$\pm$0.17 & +0.48$\pm$0.17 &  & +0.20$\pm$0.17  & +0.40$\pm$0.17 \\

$[Ni/Fe]$ (dex)        &  & +0.30$\pm$0.17 & +0.25$\pm$0.17 &  & +0.15$\pm$0.17  & +0.10 $\pm$0.17 \\

\end{tabular}
\label{tab:tab4}
\end{table}

\section{Results}

Figure~\ref{fig:fig3} shows our best-fitting synthetic spectra superimposed on the observed spectra of the target giants in both globular clusters. The derived stellar parameters and element abundances are summarized in Table~\ref{tab:tab4}. By definition:
\begin{equation}
[X/Fe] = ({log\epsilon(X) - log\epsilon(Fe)}_{star}) - ({log\epsilon(X) - log\epsilon(Fe)}_{\odot})
\end{equation}
\begin{equation}
log\epsilon(X) = log\,n_{X}/n_{H} + 12, 
\end{equation}
where $log\,n_{X}$ is the number density of element X. 

As mentioned before, the observed spectral interval does not cover the CO molecular lines, hence we can not derive C abundance. To approach this we estimate the [N/Fe], [O/Fe], [Si/Fe], [Ti/Fe], [Ni/Fe] abundances for three distinct [C/Fe] values, consistent with the range reported for 14 red giants in the Galactic bulge by \citet{ryde_et_al_2009} and \citet{ryde_et_al_2010}. As evident from the table, taking into account the uncertainties, most of the estimates for the individual giants are in agreement and only the [N/Fe] is affected by variations of [C/Fe]. In Table~\ref{tab:tab5} we present the mean [O/Fe], [Si/Fe], [Ti/Fe], [Ni/Fe] abundances for each observed giant in Mercer~5 and 2MASS\,GC02 based on three estimates per star ([Fe/H] values for each star are also listed ). For each cluster, [Fe/H] is calculated as the mean for the two giants, [O/Fe], [Si/Fe], [Ti/Fe], [Ni/Fe] are the mean of all the individual estimates. The uncertainties of the individual measurements were used to calculate the corresponding weights. The uncertainties reported in the table represent a conservative estimate, taking into account observational uncertainties, errors of calibrations, transformations and determination of the atmospheric parameters.

\begin{figure}
\centering
    \includegraphics[width=1.0\textwidth]{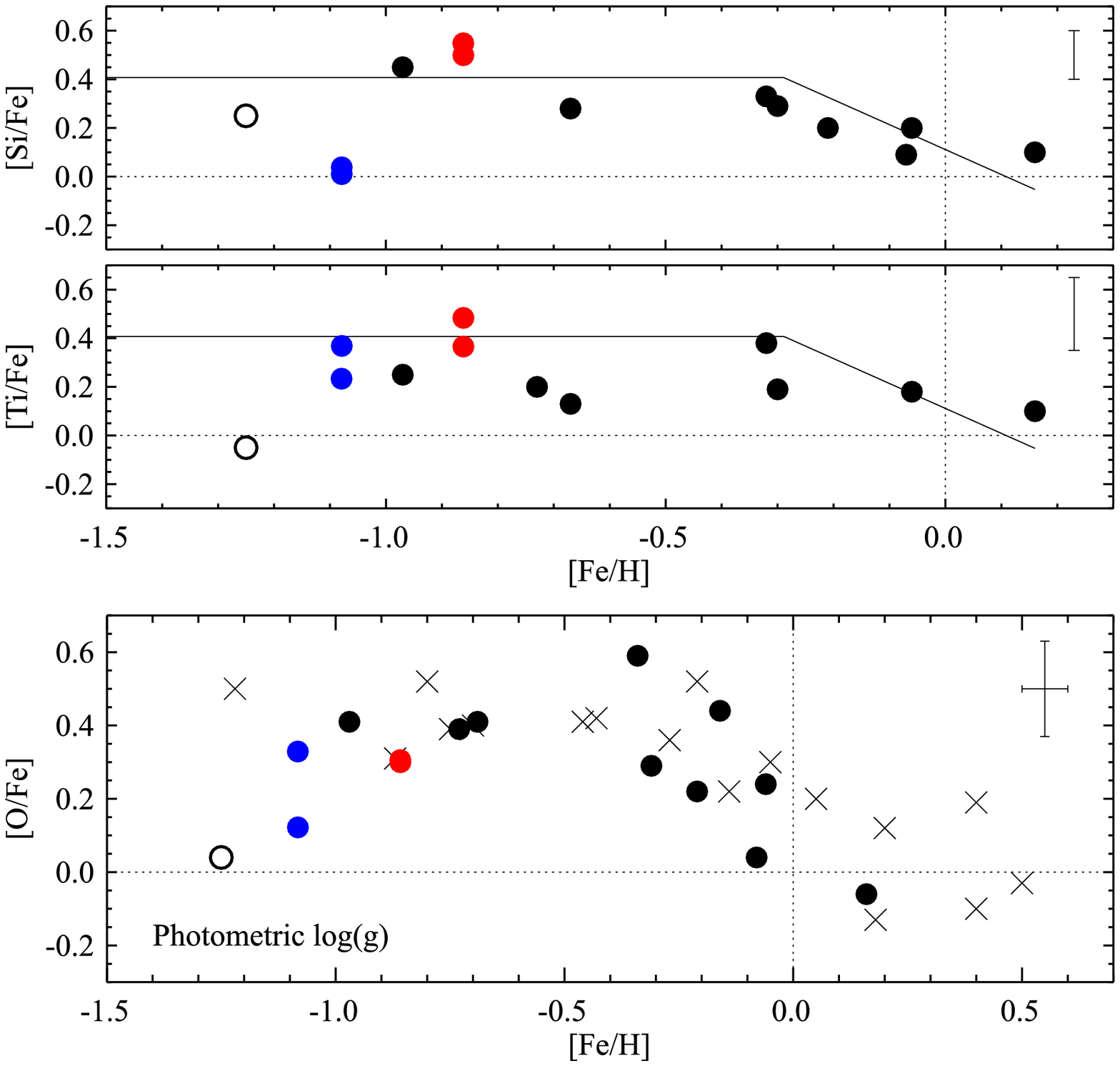}
\caption{The ratios of O, Si and Ti to Iron (top to bottom). The dark circles stand for NGC~6522 red giant stars; blue, green and pink circles are for NGC~6569, NGC~6624 and Terzan~1 red giants, the Mercer~5 and 2MASS\,GC02 red giant abundance ratios from this paper are presented as red circles and are labeled.}
\label{fig:fig4}
\end{figure}

\begin{figure}
\centering
    \includegraphics[width=1.0\textwidth]{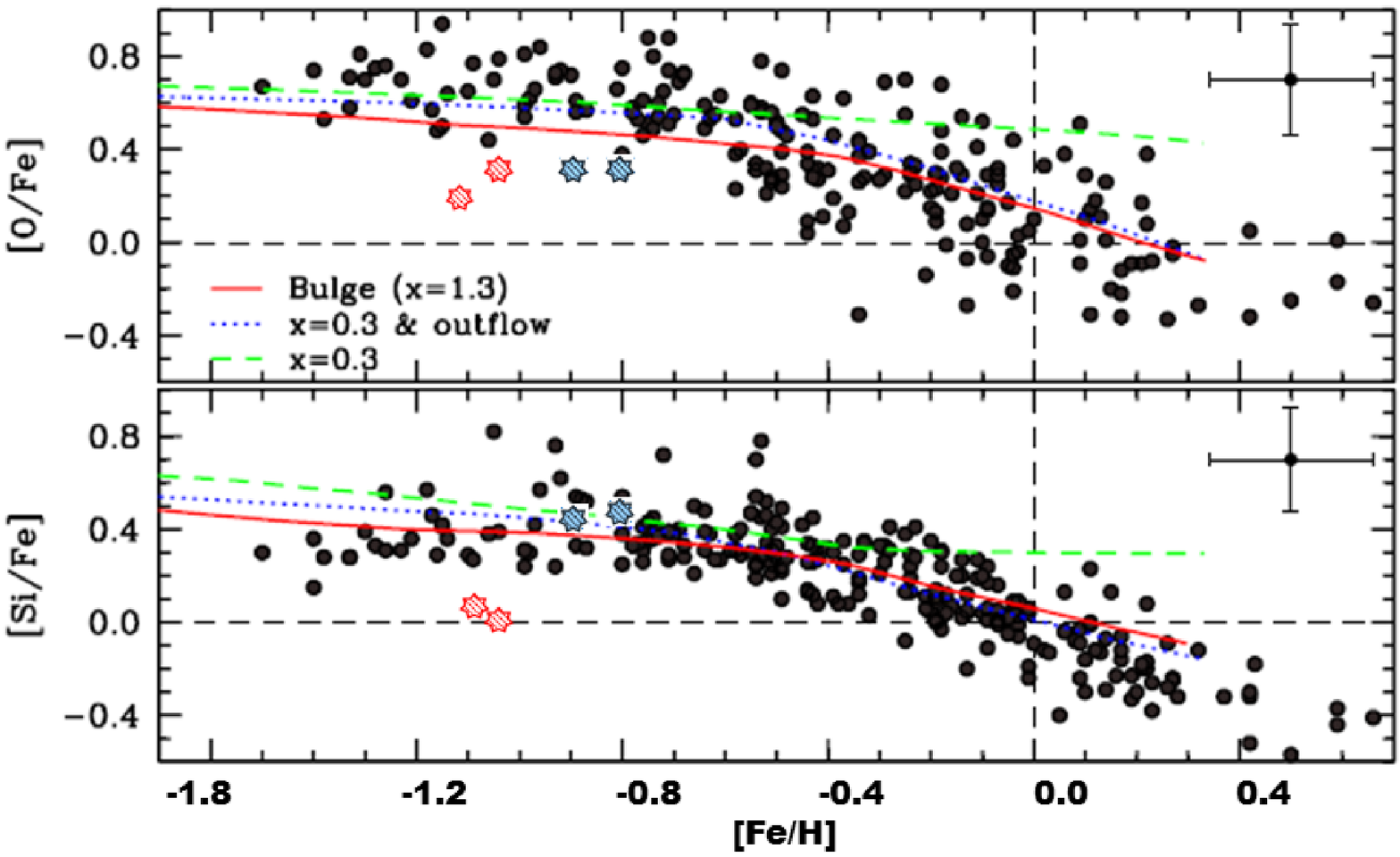}
\caption{On the Fig.~18 of \citet{johnson_et_al_2013} are overplotted  [O/Fe] and [Si/Fe] abundance rations of Mercer~5 (blue stars) and 2MASS\,GC02 (red stars) objects. The black  circles are abundance rations of 264 red giants of three combined bulge fields; the solid red line shows the predicted change in each $\alpha$-element as a function of [Fe/H], based on the bulge model of Kobayashi et al. (2011). The dashed green line is the model prediction assuming a flatter IMF (x=0.3) and the dotted blue line is the model prediction assuming x=0.3 with outflow (see \citet{johnson_et_al_2013} for details).}
\label{fig:fig5}
\end{figure}

To compare our determinations with the abundance rations measured by previous authors we selected four bulge globular clusters: NGC~6522 (\citet{barbuy_et_al_2014});  NGC~6569 and NGC~6624 (\citet{Valenti_et_al_2011}) and Terzan~1 (\citet{Valenti_et_al_2014}); as well as the abundance measurements of 264 red giant stars in three bulge fields taken from \citet{johnson_et_al_2013}. The results are shown in  Figure~\ref{fig:fig4} and Figure~\ref{fig:fig5}. 

As can be seen from the plots, the two measured stars of Mercer~5 follow the general 
trend of both bulge field and cluster stars at this metallicity. Indeed, their [O/Fe], [Si/Fe] 
and [Ti/Fe] ratios are enhanced by $\geq +$ 0.3. The 2MASS\,GC02 stars have relatively 
lower ratios, but still compatible with other bulge clusters. Therefore, abundance 
ratios alone would not allow us to confirm that these two objects are indeed
bound globular clusters. On the other hand, the bulge metallicity distribution
is populated only very sparsely at [Fe/H]$\sim$-1.0, therefore if those 4
were just bulge field stars, the probability of having all of them so metal
poor is virtually zero. Hence we confirm the cluster nature of both Mercer~5
and 2MASS\,GC02.

\begin{table}
\begin{center}
\begin{tabular}{lrrrrrr}
 \hline
 & \multicolumn{3}{c}{Mercer~5} & \multicolumn{3}{c}{2MASS\,GC02} \\[3pt]
 &  \multicolumn{1}{c}{Object \#1} & \multicolumn{1}{c}{Object \#2} & \multicolumn{1}{c}{Mean} & \multicolumn{1}{c}{Object \#1} & \multicolumn{1}{c}{Object \#4} & \multicolumn{1}{c}{Mean} \\[3pt]
\hline
[Fe/H] & $-0.90\pm0.13 $ & $-0.80\pm0.14 $ & $-0.86\pm0.14 $ & $-1.10\pm0.14 $ & $-1.05\pm0.14 $ & $-1.08\pm0.13$ \\[3pt]
[O/Fe] & $0.30\pm0.11 $ & $0.31\pm0.11 $ & $0.31\pm0.11 $ & $0.15\pm0.12 $ & $0.33\pm0.12 $ & $0.23\pm0.12$ \\[3pt]
[Si/Fe] & $0.50\pm0.14 $ & $0.55\pm0.14 $ & $0.53\pm0.14 $ & $0.04\pm0.15 $ & $0.00\pm0.15 $ & $0.02\pm0.15$ \\[3pt]
[Ti/Fe] & $0.37\pm0.17 $ & $0.48\pm0.17 $ & $0.42\pm0.17 $ & $0.23\pm0.17 $ & $0.37\pm0.17 $ & $0.30\pm0.17$ \\[3pt]
[Ni/Fe] & $0.30\pm0.17 $ & $0.25\pm0.17 $ & $0.28\pm0.17 $ & $0.17\pm0.17 $ & $0.10\pm0.17 $ & $0.13\pm0.17$ \\
\hline
\end{tabular}
\end{center}
\caption{Derived abundances for both clusters. The mean [O/Fe], [Si/Fe], [Ti/Fe], [Ni/Fe] abundances for each observed giant in Mercer~5 and 2MASS\,GC02 are based on three estimates per star, ([Fe/H] values for each star are also listed ). For each cluster, [Fe/H] is calculated as the mean for the two giants, [O/Fe],  [Si/Fe], [Ti/Fe], [Ni/Fe] are the mean of the six individual estimates.}
\label{tab:tab5}
\end{table}

\section{Summary}

We present the first chemical abundance estimates of two newly discovered Galactic globular clusters, residing in the direction of the Bulge in regions of high interstellar reddening. [Fe/H] for 2MASS\,GC02 is in agreement with earlier estimate by \citet{borissova_et_al_2007} based on moderate-resolution near-IR spectroscopy in the K  band. The metallicity for Mercer~5 is significantly higher than the value derived  by \citet{longmore_et_al_2011} using SofI/NTT moderate-resolution spectra. Based on these results we conclude that both Mercer~5 and 2MASS\,GS02 are two intermediate metal reach Bulge globular clusters, with Iron abundances of [Fe/H]=$-0.86$ and [Fe/H]=$-1.08$, respectively. The [O/Fe], [Si/Fe] and [Ti/Fe] abundance rations of Mercer~5 are enhanced by $\geq +$ 0.3, with respect to solar value, while the two observed giants of 2MASS\,GC02 show lower rations.

\section*{Acknowledgements}

Based on observations obtained at the Gemini Observatory, which is operated by the Association of Universities for Research in Astronomy, Inc., under a cooperative agreement with the NSF on behalf of the Gemini partnership: the National Science Foundation (United States), the National Research Council (Canada), CONICYT (Chile), the Australian Research Council (Australia), Minist\'{e}rio da Ci\^{e}ncia, Tecnologia e Inova\c{c}\~{a}o (Brazil) and Ministerio de Ciencia, Tecnolog\'{i}a e Innovaci\'{o}n Productiva (Argentina). This paper is based on observations obtained with the Phoenix infrared spectrograph, developed and operated by the National Optical Astronomy Observatory. The data were acquired as part of science program GS-2010A-Q-30 and 179.B-2002,VIRCAM, VISTA at ESO, Paranal Observatory. Support for JB,RK,SV,MZ is provided by the Ministry of Economy, Development, and Tourism's Millennium Science Initiative through grant IC12009, awarded to The Millennium Institute of Astrophysics (MAS) and Fondecyt Reg. No. 1120601 and No. 1110393. We acknowledge technical assistance by B. Idahl (CTIO REU 2013 program) on Figure~\ref{fig:fig1}. The authors are grateful to the anonymous referee for the constructive comments that improved the paper.

Facilities: \facility{Gemini:South(Phoenix)}

\clearpage

\begin{deluxetable}{cccccccccccccccc},
\tabletypesize{\tiny} 
\setlength{\tabcolsep}{0.005in} 
\tablewidth{0pt}
\tablecaption{Line list used to model the observed spectra. The wavelength is in air, the excitation energy is of the lower level. }
\tablehead{
\colhead{No.} &
\colhead{Wav.[$\AA$]} &
\colhead{Elem.} &
\colhead{$\chi_\mathrm{exc}$[eV]} &
\colhead{No.} &
\colhead{Wav. [$\AA$]} &
\colhead{Elem.} &
\colhead{$\chi_\mathrm{exc}$[eV]} &
\colhead{No.} &
\colhead{Wav.[$\AA$]} &
\colhead{Elem.} &
\colhead{$\chi_\mathrm{exc}$[eV]} &
\colhead{No.} &
\colhead{Wav.[$\AA$]} &
\colhead{Elem.} &
\colhead{$\chi_\mathrm{exc}$[eV]} \\}
\startdata
1	&	15500.073	&	ScI	&	4.50	&	176	&	15520.653	&	CN	&	3.76	&	351	&	15538.772	&	CN	&	3.75	&	526	&	15557.847	&	CN	&	2.65	\\
2	&	15500.241	&	VII	&	9.04	&	177	&	15520.684	&	CN	&	3.36	&	352	&	15538.799	&	CN	&	4.15	&	527	&	15558.023	&	OH	&	0.30	\\
3	&	15500.316	&	TiI	&	4.36	&	178	&	15520.855	&	CN	&	2.94	&	353	&	15538.854	&	CN	&	2.89	&	528	&	15558.045	&	CN	&	0.91	\\
4	&	15500.345	&	CN	&	0.86	&	179	&	15520.893	&	OH	&	2.19	&	354	&	15538.903	&	CN	&	2.53	&	529	&	15558.176	&	CN	&	2.66	\\
5	&	15500.650	&	MnI	&	6.20	&	180	&	15521.086	&	FeI	&	5.35	&	355	&	15539.060	&	CN	&	4.13	&	530	&	15558.585	&	TiI	&	4.50	\\
6	&	15500.708	&	CN	&	2.58	&	181	&	15521.135	&	CN	&	4.00	&	356	&	15539.126	&	OH	&	0.79	&	531	&	15558.665	&	CN	&	0.73	\\
7	&	15500.800	&	FeI	&	6.32	&	182	&	15521.207	&	CN	&	5.20	&	357	&	15539.330	&	CN	&	4.25	&	532	&	15558.880	&	CN	&	2.64	\\
8	&	15501.003	&	CN	&	0.86	&	183	&	15521.245	&	CN	&	1.72	&	358	&	15539.417	&	CrI	&	5.96	&	533	&	15558.960	&	CN	&	2.87	\\
9	&	15501.080	&	CN	&	3.15	&	184	&	15521.383	&	CN	&	2.40	&	359	&	15539.666	&	CN	&	2.53	&	534	&	15559.103	&	CN	&	2.42	\\
10	&	15501.080	&	FeI	&	5.94	&	185	&	15521.515	&	CN	&	4.41	&	360	&	15539.673	&	CN	&	4.42	&	535	&	15559.500	&	NiI	&	5.87	\\
11	&	15501.320	&	FeI	&	6.29	&	186	&	15521.690	&	FeI	&	6.32	&	361	&	15539.758	&	CN	&	3.29	&	536	&	15559.542	&	CN	&	1.01	\\
12	&	15501.345	&	CN	&	3.29	&	187	&	15522.054	&	OH	&	2.83	&	362	&	15539.777	&	CN	&	5.82	&	537	&	15559.556	&	CN	&	2.67	\\
13	&	15501.354	&	OH	&	3.10	&	188	&	15522.230	&	OH	&	3.12	&	363	&	15539.837	&	CN	&	0.88	&	538	&	15559.566	&	CN	&	1.00	\\
14	&	15501.511	&	CN	&	0.99	&	189	&	15522.236	&	OH	&	2.73	&	364	&	15539.992	&	CN	&	5.16	&	539	&	15559.648	&	CN	&	1.01	\\
15	&	15501.787	&	CN	&	2.84	&	190	&	15522.287	&	CN	&	2.94	&	365	&	15540.312	&	OH	&	3.48	&	540	&	15559.660	&	CN	&	1.00	\\
16	&	15501.833	&	CN	&	2.58	&	191	&	15522.460	&	CN	&	4.21	&	366	&	15540.463	&	CN	&	3.98	&	541	&	15559.801	&	CN	&	1.00	\\
17	&	15502.170	&	FeI	&	6.35	&	192	&	15522.600	&	CoI	&	6.21	&	367	&	15540.516	&	CN	&	0.88	&	542	&	15559.849	&	FeI	&	5.93	\\
18	&	15502.239	&	CN	&	5.50	&	193	&	15522.640	&	FeI	&	6.32	&	368	&	15540.518	&	CN	&	3.77	&	543	&	15559.922	&	CaI	&	5.17	\\
19	&	15502.261	&	CN	&	0.99	&	194	&	15522.672	&	OH	&	2.19	&	369	&	15540.898	&	CN	&	4.96	&	544	&	15559.973	&	CN	&	0.99	\\
20	&	15502.294	&	OH	&	2.94	&	195	&	15523.041	&	CN	&	5.20	&	370	&	15541.125	&	CN	&	4.04	&	545	&	15560.135	&	CN	&	2.51	\\
21	&	15502.429	&	CoI	&	3.41	&	196	&	15523.051	&	OH	&	2.73	&	371	&	15541.299	&	CN	&	3.98	&	546	&	15560.149	&	CN	&	1.01	\\
22	&	15502.434	&	CN	&	3.98	&	197	&	15523.386	&	CN	&	3.72	&	372	&	15541.516	&	CN	&	5.73	&	547	&	15560.208	&	CN	&	1.00	\\
23	&	15502.549	&	CN	&	2.84	&	198	&	15523.511	&	CN	&	1.45	&	373	&	15541.547	&	FeI	&	5.84	&	548	&	15560.244	&	OH	&	0.30	\\
24	&	15502.564	&	CN	&	2.55	&	199	&	15523.591	&	OH	&	3.19	&	374	&	15541.557	&	CN	&	3.83	&	549	&	15560.270	&	CN	&	1.01	\\
25	&	15502.576	&	CN	&	2.56	&	200	&	15523.807	&	CN	&	4.21	&	375	&	15541.644	&	OH	&	0.89	&	550	&	15560.287	&	CN	&	1.00	\\
26	&	15502.640	&	SiI	&	7.13	&	201	&	15523.909	&	CN	&	0.88	&	376	&	15541.654	&	CN	&	4.42	&	551	&	15560.324	&	CN	&	3.71	\\
27	&	15502.836	&	OH	&	3.44	&	202	&	15523.998	&	CoI	&	6.05	&	377	&	15541.818	&	CN	&	3.58	&	552	&	15560.578	&	CN	&	1.00	\\
28	&	15502.934	&	CN	&	2.56	&	203	&	15524.003	&	CN	&	3.46	&	378	&	15541.850	&	CN	&	1.26	&	553	&	15560.685	&	CN	&	2.51	\\
29	&	15503.043	&	OH	&	2.86	&	204	&	15524.277	&	NiI	&	2.74	&	379	&	15541.852	&	FeI	&	5.97	&	554	&	15560.704	&	CN	&	2.64	\\
30	&	15503.246	&	ScI	&	4.17	&	205	&	15524.300	&	FeI	&	5.79	&	380	&	15541.857	&	FeI	&	6.37	&	555	&	15560.780	&	FeI	&	6.35	\\
31	&	15503.246	&	CrI	&	3.38	&	206	&	15524.451	&	CN	&	0.88	&	381	&	15542.016	&	SiI	&	7.01	&	556	&	15561.041	&	CoI	&	6.08	\\
32	&	15503.441	&	OH	&	2.72	&	207	&	15524.543	&	FeI	&	5.79	&	382	&	15542.090	&	SiI	&	7.01	&	557	&	15561.242	&	CN	&	3.72	\\
33	&	15503.840	&	FeI	&	5.97	&	208	&	15524.615	&	CN	&	4.21	&	383	&	15542.090	&	FeI	&	5.64	&	558	&	15561.251	&	SiI	&	7.04	\\
34	&	15503.943	&	CoI	&	5.73	&	209	&	15524.822	&	CN	&	2.55	&	384	&	15542.108	&	CN	&	0.83	&	559	&	15561.268	&	FeI	&	6.71	\\
35	&	15503.967	&	VI	&	4.72	&	210	&	15524.832	&	CN	&	4.55	&	385	&	15542.146	&	OH	&	0.89	&	560	&	15561.399	&	CN	&	1.01	\\
36	&	15503.994	&	CN	&	3.37	&	211	&	15524.847	&	OH	&	0.84	&	386	&	15542.173	&	CN	&	4.13	&	561	&	15561.457	&	CN	&	1.00	\\
37	&	15504.083	&	CN	&	3.21	&	212	&	15525.227	&	FeI	&	5.84	&	387	&	15542.197	&	TiI	&	4.69	&	562	&	15561.523	&	CN	&	1.00	\\
38	&	15504.126	&	CN	&	3.72	&	213	&	15525.360	&	CN	&	5.82	&	388	&	15542.205	&	CN	&	5.20	&	563	&	15561.535	&	CN	&	1.01	\\
39	&	15504.554	&	CN	&	3.72	&	214	&	15525.406	&	CN	&	4.61	&	389	&	15542.297	&	CN	&	1.91	&	564	&	15561.748	&	CN	&	2.68	\\
40	&	15505.107	&	CN	&	3.94	&	215	&	15525.435	&	CN	&	4.67	&	390	&	15542.316	&	CN	&	5.12	&	565	&	15562.080	&	VI	&	4.63	\\
41	&	15505.326	&	OH	&	0.52	&	216	&	15525.495	&	CN	&	4.41	&	391	&	15542.611	&	CN	&	5.08	&	566	&	15562.131	&	CN	&	4.25	\\
42	&	15505.350	&	CN	&	5.21	&	217	&	15525.514	&	CN	&	2.94	&	392	&	15542.731	&	TiI	&	4.39	&	567	&	15562.143	&	CN	&	2.50	\\
43	&	15505.524	&	OH	&	1.43	&	218	&	15525.531	&	CN	&	2.72	&	393	&	15542.980	&	CN	&	1.20	&	568	&	15562.291	&	CN	&	1.15	\\
44	&	15505.526	&	CN	&	1.45	&	219	&	15525.661	&	VI	&	4.88	&	394	&	15543.326	&	CN	&	4.00	&	569	&	15562.300	&	NiI	&	6.37	\\
45	&	15505.591	&	CN	&	2.99	&	220	&	15525.734	&	TiII	&	8.10	&	395	&	15543.357	&	TiI	&	4.86	&	570	&	15562.436	&	CN	&	6.54	\\
46	&	15505.747	&	OH	&	0.52	&	221	&	15525.738	&	TiI	&	4.79	&	396	&	15543.633	&	CN	&	4.05	&	571	&	15562.441	&	CN	&	2.50	\\
47	&	15505.771	&	TiI	&	4.39	&	222	&	15525.775	&	OH	&	3.51	&	397	&	15543.780	&	TiI	&	1.88	&	572	&	15562.460	&	CN	&	6.32	\\
48	&	15505.782	&	OH	&	1.89	&	223	&	15525.775	&	CN	&	2.55	&	398	&	15543.785	&	OH	&	0.84	&	573	&	15562.601	&	OH	&	2.77	\\
49	&	15505.846	&	CN	&	4.27	&	224	&	15525.934	&	CN	&	2.51	&	399	&	15543.838	&	TiI	&	4.79	&	574	&	15563.095	&	OH	&	2.77	\\
50	&	15505.849	&	OH	&	1.43	&	225	&	15525.963	&	CN	&	4.21	&	400	&	15543.846	&	CN	&	3.58	&	575	&	15563.136	&	CN	&	3.58	\\
51	&	15506.052	&	OH	&	2.85	&	226	&	15526.062	&	CN	&	2.84	&	401	&	15544.152	&	CuI	&	6.79	&	576	&	15563.139	&	OH	&	2.75	\\
52	&	15506.079	&	CN	&	2.44	&	227	&	15526.083	&	CN	&	5.15	&	402	&	15544.355	&	TiI	&	2.49	&	577	&	15563.165	&	CN	&	4.29	\\
53	&	15506.099	&	OH	&	1.43	&	228	&	15526.404	&	CN	&	1.00	&	403	&	15544.452	&	CN	&	2.72	&	578	&	15563.303	&	CN	&	1.00	\\
54	&	15506.105	&	FeI	&	5.52	&	229	&	15526.414	&	CN	&	3.77	&	404	&	15544.501	&	CN	&	1.15	&	579	&	15563.306	&	CN	&	1.02	\\
55	&	15506.246	&	OH	&	1.43	&	230	&	15526.604	&	CN	&	2.45	&	405	&	15544.680	&	CN	&	2.79	&	580	&	15563.315	&	CN	&	2.63	\\
56	&	15506.246	&	OH	&	1.89	&	231	&	15526.819	&	CN	&	5.39	&	406	&	15544.730	&	CN	&	2.86	&	581	&	15563.354	&	CN	&	1.00	\\
57	&	15506.252	&	ScI	&	4.97	&	232	&	15526.841	&	CN	&	3.84	&	407	&	15544.771	&	CN	&	4.00	&	582	&	15563.376	&	CN	&	1.15	\\
58	&	15506.363	&	CN	&	4.26	&	233	&	15526.865	&	CN	&	4.55	&	408	&	15544.899	&	CN	&	4.47	&	583	&	15563.456	&	CN	&	1.02	\\
59	&	15506.408	&	CN	&	4.91	&	234	&	15526.976	&	CN	&	2.72	&	409	&	15544.948	&	CN	&	2.51	&	584	&	15563.463	&	CrI	&	5.24	\\
60	&	15506.685	&	CN	&	2.71	&	235	&	15527.043	&	CN	&	3.15	&	410	&	15545.047	&	CN	&	4.41	&	585	&	15563.778	&	CN	&	2.46	\\
61	&	15506.779	&	CN	&	1.45	&	236	&	15527.210	&	FeI	&	6.32	&	411	&	15545.332	&	CN	&	2.72	&	586	&	15563.902	&	CN	&	2.79	\\
62	&	15506.901	&	CN	&	6.39	&	237	&	15527.323	&	CN	&	3.90	&	412	&	15545.409	&	CN	&	5.63	&	587	&	15564.020	&	FeII	&	9.05	\\
63	&	15506.969	&	CN	&	5.52	&	238	&	15527.325	&	CN	&	3.84	&	413	&	15545.511	&	CN	&	2.86	&	588	&	15564.185	&	CN	&	4.25	\\
64	&	15506.980	&	SiI	&	6.73	&	239	&	15527.465	&	CN	&	2.57	&	414	&	15545.584	&	CN	&	1.15	&	589	&	15564.268	&	CN	&	2.75	\\
65	&	15507.022	&	VI	&	4.87	&	240	&	15527.470	&	CN	&	2.94	&	415	&	15545.668	&	CN	&	5.82	&	590	&	15564.369	&	FeI	&	5.61	\\
66	&	15507.043	&	PI	&	8.23	&	241	&	15527.513	&	CN	&	3.94	&	416	&	15545.782	&	CN	&	1.22	&	591	&	15564.684	&	CN	&	2.73	\\
67	&	15507.046	&	CN	&	1.00	&	242	&	15527.535	&	SiI	&	7.14	&	417	&	15546.081	&	TiI	&	4.41	&	592	&	15564.723	&	ScI	&	4.53	\\
68	&	15507.046	&	CN	&	3.21	&	243	&	15527.564	&	CN	&	4.61	&	418	&	15546.089	&	CN	&	2.60	&	593	&	15564.769	&	CN	&	2.69	\\
69	&	15507.048	&	CN	&	2.99	&	244	&	15527.629	&	CN	&	4.55	&	419	&	15546.488	&	CN	&	2.60	&	594	&	15564.793	&	CN	&	2.75	\\
70	&	15507.103	&	TiI	&	4.77	&	245	&	15527.713	&	OH	&	2.89	&	420	&	15546.531	&	CN	&	4.47	&	595	&	15564.938	&	OH	&	0.78	\\
71	&	15507.118	&	FeII	&	8.94	&	246	&	15527.837	&	CN	&	2.57	&	421	&	15546.709	&	OH	&	3.48	&	596	&	15565.230	&	FeI	&	6.32	\\
72	&	15507.156	&	CN	&	0.89	&	247	&	15527.890	&	CN	&	4.42	&	422	&	15546.780	&	CN	&	2.83	&	597	&	15565.256	&	CN	&	2.58	\\
73	&	15507.180	&	CN	&	2.71	&	248	&	15528.109	&	FeI	&	5.95	&	423	&	15546.790	&	CN	&	4.47	&	598	&	15565.336	&	CN	&	2.42	\\
74	&	15507.241	&	CN	&	0.71	&	249	&	15528.121	&	OH	&	1.35	&	424	&	15546.818	&	OH	&	2.75	&	599	&	15565.356	&	CN	&	0.91	\\
75	&	15507.332	&	CN	&	5.21	&	250	&	15528.128	&	CN	&	4.92	&	425	&	15546.838	&	CN	&	1.22	&	600	&	15565.390	&	CN	&	4.05	\\
76	&	15507.500	&	CN	&	4.49	&	251	&	15528.160	&	CN	&	2.96	&	426	&	15546.848	&	CN	&	5.44	&	601	&	15565.448	&	CN	&	1.76	\\
77	&	15507.623	&	TiI	&	5.24	&	252	&	15528.218	&	CN	&	1.00	&	427	&	15546.920	&	CN	&	1.55	&	602	&	15565.588	&	CN	&	5.35	\\
78	&	15507.816	&	CN	&	2.38	&	253	&	15528.365	&	CN	&	4.88	&	428	&	15547.041	&	CN	&	4.21	&	603	&	15565.644	&	CN	&	2.79	\\
79	&	15507.844	&	CN	&	0.89	&	254	&	15528.575	&	CN	&	2.72	&	429	&	15547.940	&	OH	&	1.56	&	604	&	15565.734	&	CN	&	0.99	\\
80	&	15508.075	&	CN	&	2.64	&	255	&	15528.577	&	CN	&	4.35	&	430	&	15548.190	&	CN	&	0.99	&	605	&	15565.770	&	CN	&	0.99	\\
81	&	15508.385	&	TiI	&	4.77	&	256	&	15528.589	&	CN	&	2.67	&	431	&	15548.349	&	CN	&	3.67	&	606	&	15565.817	&	OH	&	0.90	\\
82	&	15508.393	&	CN	&	4.80	&	257	&	15528.599	&	CN	&	4.67	&	432	&	15548.422	&	CN	&	4.47	&	607	&	15565.838	&	OH	&	3.66	\\
83	&	15508.477	&	CN	&	1.22	&	258	&	15528.618	&	CN	&	3.77	&	433	&	15548.480	&	CN	&	3.89	&	608	&	15565.879	&	CN	&	1.02	\\
84	&	15508.670	&	OH	&	1.89	&	259	&	15528.631	&	OH	&	3.06	&	434	&	15548.487	&	OH	&	3.66	&	609	&	15565.886	&	VI	&	4.64	\\
85	&	15508.677	&	CN	&	6.07	&	260	&	15528.670	&	OH	&	3.21	&	435	&	15548.673	&	CN	&	2.87	&	610	&	15565.962	&	OH	&	2.78	\\
86	&	15508.718	&	CN	&	2.64	&	261	&	15528.768	&	CN	&	3.94	&	436	&	15548.908	&	CN	&	2.61	&	611	&	15565.996	&	OH	&	0.90	\\
87	&	15508.799	&	CN	&	0.97	&	262	&	15528.891	&	CN	&	4.41	&	437	&	15548.914	&	OH	&	0.93	&	612	&	15566.044	&	CN	&	1.02	\\
88	&	15509.016	&	CN	&	3.46	&	263	&	15528.903	&	CN	&	2.94	&	438	&	15548.978	&	CaI	&	5.18	&	613	&	15566.051	&	CN	&	5.16	\\
89	&	15509.507	&	CN	&	2.48	&	264	&	15528.915	&	CN	&	3.15	&	439	&	15549.206	&	CN	&	3.19	&	614	&	15566.255	&	CN	&	4.12	\\
90	&	15509.526	&	CN	&	0.97	&	265	&	15528.994	&	CN	&	1.00	&	440	&	15549.501	&	CN	&	3.90	&	615	&	15566.274	&	CN	&	6.40	\\
91	&	15509.779	&	CN	&	2.48	&	266	&	15528.994	&	CN	&	3.06	&	441	&	15549.541	&	OH	&	2.59	&	616	&	15566.393	&	CN	&	4.96	\\
92	&	15509.863	&	CN	&	2.70	&	267	&	15529.105	&	CN	&	4.92	&	442	&	15549.554	&	CN	&	2.83	&	617	&	15566.603	&	OH	&	2.68	\\
93	&	15510.178	&	CN	&	4.20	&	268	&	15529.168	&	CN	&	2.96	&	443	&	15549.736	&	OH	&	2.18	&	618	&	15566.703	&	CN	&	2.63	\\
94	&	15510.358	&	OH	&	1.89	&	269	&	15529.226	&	CN	&	5.84	&	444	&	15549.786	&	CN	&	3.90	&	619	&	15566.725	&	FeI	&	6.35	\\
95	&	15510.648	&	OH	&	0.26	&	270	&	15529.243	&	CN	&	2.77	&	445	&	15549.803	&	CN	&	4.03	&	620	&	15566.907	&	CN	&	4.50	\\
96	&	15510.717	&	CN	&	3.46	&	271	&	15529.316	&	OH	&	3.06	&	446	&	15549.902	&	OH	&	0.73	&	621	&	15566.941	&	CN	&	3.84	\\
97	&	15510.847	&	CN	&	2.72	&	272	&	15529.344	&	CN	&	4.61	&	447	&	15549.948	&	CN	&	2.87	&	622	&	15567.001	&	ScI	&	4.97	\\
98	&	15511.088	&	CN	&	4.20	&	273	&	15529.453	&	CN	&	2.72	&	448	&	15550.226	&	CN	&	4.10	&	623	&	15567.014	&	CN	&	3.94	\\
99	&	15511.117	&	SiI	&	7.17	&	274	&	15529.657	&	OH	&	3.48	&	449	&	15550.320	&	CN	&	2.66	&	624	&	15567.188	&	SiI	&	7.11	\\
100	&	15511.528	&	FeI	&	5.48	&	275	&	15529.662	&	CN	&	4.55	&	450	&	15550.357	&	CN	&	4.03	&	625	&	15567.261	&	FeI	&	6.35	\\
101	&	15511.665	&	CN	&	5.73	&	276	&	15529.773	&	CN	&	3.15	&	451	&	15550.381	&	CN	&	0.89	&	626	&	15567.530	&	CN	&	6.42	\\
102	&	15511.769	&	CN	&	5.52	&	277	&	15529.846	&	CN	&	4.67	&	452	&	15550.450	&	FeI	&	6.34	&	627	&	15567.552	&	OH	&	3.00	\\
103	&	15511.810	&	CN	&	0.88	&	278	&	15529.913	&	CN	&	1.25	&	453	&	15550.553	&	CN	&	2.65	&	628	&	15567.575	&	OH	&	0.73	\\
104	&	15512.147	&	CN	&	3.62	&	279	&	15529.928	&	CN	&	2.67	&	454	&	15550.560	&	FeI	&	6.11	&	629	&	15567.680	&	VI	&	4.62	\\
105	&	15512.291	&	CN	&	2.37	&	280	&	15530.010	&	CN	&	2.64	&	455	&	15550.623	&	CN	&	3.19	&	630	&	15567.704	&	CN	&	5.16	\\
106	&	15512.575	&	OH	&	0.86	&	281	&	15530.043	&	FeI	&	3.57	&	456	&	15550.645	&	CN	&	2.96	&	631	&	15567.724	&	CN	&	3.84	\\
107	&	15512.635	&	CN	&	4.60	&	282	&	15530.080	&	CN	&	0.89	&	457	&	15550.867	&	CN	&	3.90	&	632	&	15567.728	&	VI	&	4.59	\\
108	&	15512.724	&	VI	&	4.68	&	283	&	15530.162	&	CN	&	2.65	&	458	&	15550.949	&	CN	&	0.89	&	633	&	15567.869	&	CN	&	2.55	\\
109	&	15512.761	&	CN	&	2.48	&	284	&	15530.186	&	CN	&	2.51	&	459	&	15550.964	&	CN	&	3.06	&	634	&	15568.180	&	CN	&	1.29	\\
110	&	15512.771	&	OH	&	0.86	&	285	&	15530.195	&	VI	&	5.45	&	460	&	15550.981	&	CN	&	2.66	&	635	&	15568.187	&	SiI	&	7.11	\\
111	&	15512.891	&	CN	&	0.82	&	286	&	15530.205	&	CN	&	3.06	&	461	&	15551.172	&	CN	&	5.44	&	636	&	15568.325	&	FeI	&	5.88	\\
112	&	15513.146	&	CN	&	5.31	&	287	&	15530.316	&	CN	&	2.77	&	462	&	15551.430	&	FeI	&	6.35	&	637	&	15568.383	&	VI	&	4.67	\\
113	&	15513.208	&	CN	&	1.12	&	288	&	15530.654	&	CN	&	2.64	&	463	&	15551.636	&	CN	&	2.89	&	638	&	15568.567	&	CN	&	4.96	\\
114	&	15513.336	&	CN	&	2.67	&	289	&	15530.683	&	CN	&	4.30	&	464	&	15551.714	&	CN	&	4.14	&	639	&	15568.614	&	CN	&	2.69	\\
115	&	15513.367	&	CN	&	3.54	&	290	&	15530.810	&	CN	&	0.89	&	465	&	15551.818	&	VI	&	4.66	&	640	&	15568.650	&	CN	&	2.45	\\
116	&	15513.468	&	OH	&	0.92	&	291	&	15531.103	&	OH	&	3.29	&	466	&	15551.861	&	CN	&	2.65	&	641	&	15568.744	&	CN	&	0.99	\\
117	&	15513.477	&	OH	&	0.26	&	292	&	15531.124	&	OH	&	2.94	&	467	&	15552.108	&	TiII	&	8.11	&	642	&	15568.764	&	CN	&	0.99	\\
118	&	15513.511	&	TiI	&	4.51	&	293	&	15531.280	&	TiI	&	4.65	&	468	&	15552.222	&	FeI	&	5.62	&	643	&	15568.780	&	OH	&	0.30	\\
119	&	15513.675	&	CN	&	1.06	&	294	&	15531.494	&	CN	&	4.84	&	469	&	15552.254	&	OH	&	2.74	&	644	&	15568.853	&	OH	&	3.29	\\
120	&	15513.800	&	OH	&	0.92	&	295	&	15531.503	&	CN	&	4.61	&	470	&	15552.268	&	CN	&	3.89	&	645	&	15568.955	&	CN	&	2.86	\\
121	&	15514.195	&	CN	&	2.53	&	296	&	15531.646	&	CN	&	3.15	&	471	&	15552.747	&	CN	&	0.90	&	646	&	15568.980	&	CN	&	4.50	\\
122	&	15514.270	&	CN	&	1.12	&	297	&	15531.713	&	CN	&	3.12	&	472	&	15552.769	&	CN	&	3.89	&	647	&	15569.103	&	SiI	&	7.11	\\
123	&	15514.280	&	FeI	&	6.29	&	298	&	15531.750	&	FeI	&	5.64	&	473	&	15552.885	&	CN	&	4.14	&	648	&	15569.135	&	CN	&	1.03	\\
124	&	15514.427	&	OH	&	3.12	&	299	&	15532.116	&	CN	&	4.35	&	474	&	15553.245	&	VII	&	5.54	&	649	&	15569.240	&	FeI	&	5.51	\\
125	&	15514.496	&	CN	&	1.06	&	300	&	15532.263	&	SiI	&	7.14	&	475	&	15553.313	&	CN	&	4.14	&	650	&	15569.314	&	CN	&	1.03	\\
126	&	15514.691	&	SiI	&	7.09	&	301	&	15532.449	&	SiI	&	6.72	&	476	&	15553.340	&	CN	&	2.58	&	651	&	15569.486	&	CN	&	4.21	\\
127	&	15514.802	&	CN	&	1.33	&	302	&	15532.536	&	CN	&	4.29	&	477	&	15553.560	&	FeI	&	5.48	&	652	&	15569.490	&	OH	&	0.84	\\
128	&	15514.891	&	CN	&	1.52	&	303	&	15532.802	&	OH	&	3.48	&	478	&	15553.577	&	CN	&	2.89	&	653	&	15569.576	&	CN	&	4.74	\\
129	&	15515.117	&	OH	&	2.21	&	304	&	15533.347	&	CN	&	4.29	&	479	&	15553.659	&	CN	&	1.08	&	654	&	15569.741	&	CoI	&	5.75	\\
130	&	15515.368	&	TiI	&	2.31	&	305	&	15533.389	&	OH	&	1.35	&	480	&	15553.727	&	CN	&	2.58	&	655	&	15569.903	&	CN	&	2.61	\\
131	&	15515.373	&	VI	&	4.65	&	306	&	15533.710	&	OH	&	2.33	&	481	&	15554.146	&	CN	&	4.41	&	656	&	15569.911	&	CN	&	2.40	\\
132	&	15515.416	&	CN	&	2.38	&	307	&	15533.849	&	CN	&	0.99	&	482	&	15554.446	&	OH	&	0.79	&	657	&	15569.984	&	OH	&	2.78	\\
133	&	15515.445	&	OH	&	2.18	&	308	&	15533.977	&	SiI	&	7.14	&	483	&	15554.501	&	CN	&	1.08	&	658	&	15570.044	&	CN	&	3.81	\\
134	&	15515.617	&	OH	&	2.18	&	309	&	15534.020	&	CN	&	3.67	&	484	&	15554.510	&	FeI	&	6.28	&	659	&	15570.073	&	CN	&	3.70	\\
135	&	15515.768	&	FeI	&	6.29	&	310	&	15534.081	&	CN	&	4.29	&	484	&	15554.510	&	fEi	&	6.28	&	660	&	15570.202	&	NiII	&	8.42	\\
136	&	15515.777	&	CN	&	2.50	&	311	&	15534.182	&	CN	&	3.36	&	486	&	15554.557	&	TiI	&	4.43	&	661	&	15570.306	&	CN	&	3.10	\\
137	&	15515.797	&	CN	&	1.33	&	312	&	15534.260	&	FeI	&	5.64	&	487	&	15554.603	&	CN	&	1.28	&	662	&	15570.575	&	CN	&	3.41	\\
138	&	15515.876	&	TiI	&	4.79	&	313	&	15534.306	&	OH	&	0.84	&	488	&	15554.625	&	VI	&	4.61	&	663	&	15570.752	&	CN	&	5.08	\\
139	&	15515.931	&	CN	&	1.18	&	314	&	15534.368	&	CN	&	3.68	&	489	&	15554.697	&	VII	&	5.87	&	664	&	15570.861	&	CN	&	2.62	\\
140	&	15515.969	&	OH	&	3.06	&	315	&	15534.399	&	CN	&	2.45	&	490	&	15554.855	&	CN	&	5.01	&	665	&	15571.055	&	CN	&	2.61	\\
141	&	15516.151	&	CN	&	4.82	&	316	&	15534.455	&	CN	&	2.83	&	491	&	15554.939	&	OH	&	2.75	&	666	&	15571.099	&	VI	&	4.68	\\
142	&	15516.284	&	OH	&	2.83	&	317	&	15534.602	&	CN	&	0.99	&	492	&	15555.112	&	OH	&	3.66	&	667	&	15571.120	&	FeI	&	5.88	\\
143	&	15516.418	&	CN	&	1.00	&	318	&	15534.665	&	CN	&	4.13	&	493	&	15555.120	&	NiI	&	5.28	&	668	&	15571.152	&	CN	&	1.22	\\
144	&	15516.537	&	OH	&	3.29	&	319	&	15534.892	&	CN	&	4.29	&	494	&	15555.138	&	CN	&	4.50	&	669	&	15571.220	&	CN	&	3.19	\\
145	&	15516.660	&	OH	&	2.89	&	320	&	15534.986	&	CN	&	2.73	&	495	&	15555.210	&	NiI	&	5.28	&	670	&	15571.417	&	CN	&	4.21	\\
146	&	15516.720	&	FeI	&	6.29	&	321	&	15535.167	&	CN	&	1.33	&	496	&	15555.370	&	NiI	&	5.49	&	671	&	15571.511	&	CN	&	5.35	\\
147	&	15517.095	&	CN	&	2.77	&	322	&	15535.182	&	CN	&	4.27	&	497	&	15555.641	&	ScI	&	5.06	&	672	&	15571.664	&	CN	&	5.41	\\
148	&	15517.228	&	OH	&	2.85	&	323	&	15535.317	&	CN	&	2.49	&	498	&	15555.700	&	CN	&	1.02	&	673	&	15571.729	&	VI	&	2.58	\\
149	&	15517.275	&	CoI	&	5.74	&	324	&	15535.329	&	CN	&	3.91	&	499	&	15555.720	&	CN	&	1.36	&	674	&	15571.740	&	FeI	&	6.32	\\
150	&	15517.367	&	CN	&	1.20	&	325	&	15535.353	&	CN	&	5.50	&	500	&	15555.750	&	CN	&	1.49	&	675	&	15571.822	&	CN	&	2.66	\\
151	&	15517.487	&	CN	&	3.29	&	326	&	15535.462	&	OH	&	0.51	&	501	&	15555.765	&	OH	&	1.50	&	676	&	15571.834	&	CN	&	3.82	\\
152	&	15517.815	&	CN	&	4.35	&	327	&	15535.498	&	CN	&	2.73	&	502	&	15555.835	&	CN	&	4.14	&	677	&	15571.897	&	CN	&	3.10	\\
153	&	15518.135	&	CN	&	2.49	&	328	&	15535.602	&	CN	&	2.49	&	503	&	15556.016	&	NiI	&	5.28	&	678	&	15572.084	&	OH	&	0.30	\\
154	&	15518.166	&	CN	&	2.77	&	329	&	15535.816	&	CN	&	1.06	&	504	&	15556.058	&	CN	&	4.04	&	679	&	15572.166	&	ScI	&	5.54	\\
155	&	15518.289	&	CN	&	2.58	&	330	&	15535.829	&	CN	&	0.90	&	505	&	15556.106	&	OH	&	1.50	&	680	&	15572.217	&	CN	&	5.08	\\
156	&	15518.395	&	CN	&	1.20	&	331	&	15536.224	&	CN	&	4.92	&	506	&	15556.115	&	OH	&	1.49	&	681	&	15572.230	&	OH	&	3.19	\\
157	&	15518.670	&	CN	&	2.49	&	332	&	15536.642	&	CN	&	1.06	&	507	&	15556.122	&	CuI	&	6.55	&	682	&	15572.312	&	CN	&	0.84	\\
158	&	15518.675	&	CN	&	2.58	&	333	&	15536.706	&	OH	&	0.51	&	508	&	15556.384	&	CN	&	4.30	&	683	&	15572.334	&	CN	&	0.99	\\
159	&	15518.708	&	CN	&	4.14	&	334	&	15536.773	&	CN	&	2.71	&	509	&	15556.593	&	CN	&	1.02	&	684	&	15572.484	&	CN	&	2.66	\\
160	&	15518.720	&	ScI	&	5.10	&	335	&	15536.895	&	OH	&	3.44	&	510	&	15556.670	&	FeI	&	5.93	&	685	&	15572.632	&	TiI	&	4.67	\\
161	&	15518.764	&	CN	&	4.04	&	336	&	15536.997	&	CN	&	0.99	&	511	&	15556.711	&	TiI	&	5.30	&	686	&	15572.651	&	VI	&	4.65	\\
162	&	15518.793	&	CN	&	1.23	&	337	&	15537.181	&	OH	&	0.84	&	512	&	15556.919	&	CN	&	4.04	&	687	&	15573.083	&	CN	&	1.03	\\
163	&	15518.829	&	CN	&	5.07	&	338	&	15537.207	&	OH	&	0.78	&	513	&	15556.939	&	CN	&	1.36	&	688	&	15573.261	&	CN	&	1.49	\\
164	&	15518.900	&	FeI	&	6.28	&	339	&	15537.227	&	CN	&	3.67	&	514	&	15556.962	&	OH	&	1.50	&	689	&	15573.277	&	CN	&	1.03	\\
165	&	15519.065	&	CN	&	3.29	&	340	&	15537.253	&	CN	&	3.27	&	515	&	15557.006	&	CN	&	3.74	&	690	&	15573.294	&	CN	&	2.70	\\
166	&	15519.084	&	CN	&	3.36	&	341	&	15537.410	&	OH	&	0.48	&	516	&	15557.006	&	CN	&	4.14	&	691	&	15573.466	&	CN	&	3.90	\\
167	&	15519.100	&	FeI	&	6.29	&	342	&	15537.450	&	FeI	&	5.79	&	517	&	15557.212	&	OH	&	1.57	&	692	&	15573.673	&	CN	&	2.53	\\
168	&	15519.360	&	FeI	&	6.29	&	343	&	15537.572	&	FeI	&	5.79	&	518	&	15557.387	&	CN	&	1.49	&	693	&	15573.724	&	CN	&	5.50	\\
169	&	15519.600	&	CN	&	3.61	&	344	&	15537.690	&	FeI	&	6.32	&	519	&	15557.447	&	CN	&	3.74	&	694	&	15573.821	&	CN	&	4.68	\\
170	&	15519.636	&	VI	&	4.11	&	345	&	15537.777	&	CN	&	4.13	&	520	&	15557.602	&	VI	&	4.68	&	695	&	15573.976	&	CrI	&	5.94	\\
171	&	15519.942	&	FeI	&	2.48	&	346	&	15538.060	&	CoI	&	5.74	&	521	&	15557.607	&	CN	&	2.66	&	696	&	15574.060	&	FeI	&	6.31	\\
172	&	15519.949	&	ScI	&	3.81	&	347	&	15538.081	&	CN	&	0.76	&	522	&	15557.684	&	CN	&	2.87	&	697	&	15574.599	&	CN	&	2.42	\\
173	&	15520.115	&	SiI	&	7.11	&	348	&	15538.084	&	CN	&	5.20	&	523	&	15557.689	&	TiI	&	5.28	&	698	&	15574.667	&	CN	&	5.85	\\
174	&	15520.154	&	CN	&	5.37	&	349	&	15538.434	&	OH	&	0.48	&	524	&	15557.735	&	CN	&	2.96	&	699	&	15574.837	&	CN	&	2.61	\\
175	&	15520.262	&	CN	&	4.41	&	350	&	15538.463	&	SiI	&	6.76	&	525	&	15557.790	&	SiI	&	5.96	&		&		&		&		\\
\enddata
\label{line_list}
\end{deluxetable}

\end{document}